\documentclass[journal]{vgtc}                     

\onlineid{1034}

\vgtccategory{Research}

\title{UniScale: Exploring Unimanual Gesture Mapping Strategies for Gaze+Pinch-based Scaling Interaction}

\author{%
  \authororcid{Kyoungwhan Mheen}{0009-0007-9282-0717},
  \authororcid{Jinwook Kim}{0000-0002-1962-5815},
  \authororcid{Sang Ho Yoon}{0000-0002-3780-5350}
}
\authorfooter{
  \item
  	Kyoungwhan Mheen is with the Graduate School of Culture Technology at KAIST. E-mail: kwmheen@kaist.ac.kr.
  \item
  	Jinwook Kim (Corresponding author) is with the Graduate School of Culture Technology at KAIST. E-mail: jinwook.kim31@kaist.ac.kr.
  \item
  	Sang Ho Yoon (Corresponding author) is with the Graduate School of Culture Technology at KAIST. E-mail: sangho@kaist.ac.kr. 
}

\abstract{%
  Object scaling serves as a fundamental spatial manipulation that enables complex and productive tasks in XR environments. This paper investigates unimanual scaling techniques for XR using gaze and hand interactions. We propose UniScale, a set of unimanual alternatives to the standard bimanual pinch, allowing users to scale objects while preserving hand availability for concurrent spatial manipulations. We design five distinct mapping strategies based on physical metaphors, exploring unimanual control that varies depth, angle, micro-gestures, and finger-distance input. We then compare these techniques against a standard bimanual baseline, in which users adjust the inter-hand distance via a bimanual pinch gesture. In a user study, we evaluate their effectiveness in a 3D object scaling task under both clutching and clutching-free conditions. The results indicate that while bimanual scaling relies on clutching for stable control, unimanual techniques excel in clutching-free conditions, significantly reducing physical hand movement. From the results, we derive valuable design implications for developing efficient 3D multimodal interactions in XR.
}


\teaser{
  \centering
  \includegraphics[width=\linewidth, alt={A view of clouds with orange sunrays shining through from behind.}]{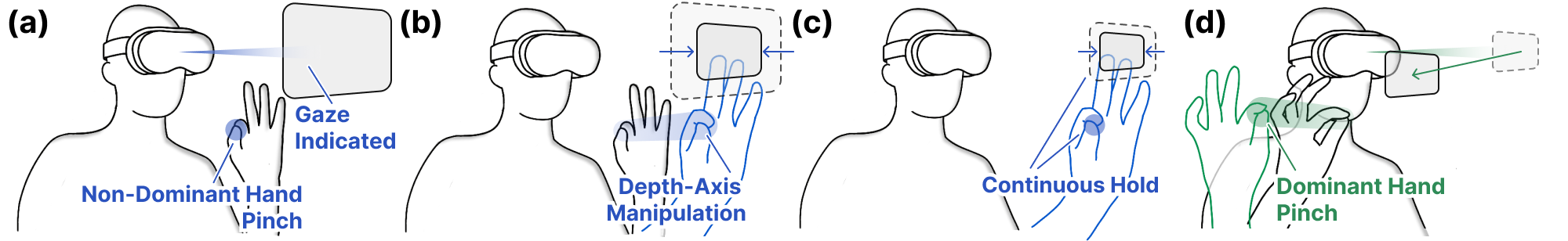}
  \caption{%
  	UniScale interaction (uniDepth) workflow for XR object manipulation. (a) Gaze indicates the target, and a pinch with the non-dominant hand activates the interaction. (b) Moving forward or backward along the depth axis to manipulate the object's scale. (c) Under the clutching-free condition, UniScale promotes continuous and uninterrupted scaling by allowing users to maintain the pinch without deactivation. (d) Complementary interactions like translation are assigned to the dominant hand pinch, decoupling them from the scaling workflow and enabling simultaneous operation.%
  }
  \label{fig:teaser}
}

\usepackage{booktabs}                  
\usepackage{lipsum}                    
\usepackage{mwe}                       
\usepackage{ccicons}                   

\usepackage{mathptmx}                  

\begin{document}


\firstsection{Introduction}

\maketitle

The integration of advanced sensors into modern extended reality (XR) head-mounted displays (HMDs) has driven the development and commercialization of natural interaction techniques~\cite{kim2025t2iray, kin2024stmg}. Among these, the Gaze+Pinch interaction model~\cite{pfeuffer2017gaze+} is increasingly becoming standardized as the primary interaction technique due to its intuitive targeting and ergonomic efficiency~\cite{gonzalez2024guidelines, pfeuffer2024design}. Despite its widespread adoption, however, recent research has predominantly focused on foundational selection mechanics~\cite{wagner2023fitts}.

Spatial manipulation, encompassing rotation, scaling, and translation (RST), plays a vital role in executing complex and productive spatial tasks in XR environments~\cite{lee2025scaling, yu2024accurate, wagner2025study, drey2023investigating}. Despite its critical importance, there remains a notable lack of RST interaction functions that seamlessly integrate with the gaze-and-pinch paradigm~\cite{turner2015gaze+}. Therefore, it requires intuitive, affordance-driven interaction techniques that naturally guide user actions while remaining clearly distinguishable from, yet fluidly integrated with, foundational gaze-and-pinch selection~\cite{vieira2024understanding}. Such advancements are essential for XR to achieve ubiquitous adoption across both daily use and advanced spatial computing tasks.

In our study, we focus specifically on the scaling task to bridge this gap. While Gaze+Pinch excels at rapid and unimanual target acquisition, subsequent scaling tasks often force users to switch to a bimanual symmetric pinch that disrupts the single-handed workflow~\cite{song2012handle}. Although this two-handed technique has become the de facto standard due to its familiar mobile metaphor and isomorphic size mapping (i.e., a 1:1 correspondence between the physical user action and the resulting virtual changes~\cite{laviola20173d}), its structural limitations severely hinder advanced XR interactions~\cite{avery2014pinch}. Specifically, occupying both hands simultaneously reduces the efficiency of Gaze+Pinch and interrupts the fluid transition to subsequent spatial manipulations, such as translating or rotating objects. Moreover, the repetitive clutching required for extensive scaling maneuvers accelerates physical fatigue. Coupled with the absence of a real-world physical metaphor, these constraints highlight a largely unexplored design space for alternative scaling techniques that preserve hand availability to support more continuous and seamless spatial manipulation workflows.

To address the functional constraints of bimanual occupation and physical fatigue, we present \textit{UniScale} (Figure~\ref{fig:teaser}). A set of unimanual scaling techniques for XR that leverages indirect input via the Gaze+Pinch interaction model~\cite{pfeuffer2017gaze+}. By designing an unimanual approach and integrating intuitive physical metaphors, UniScale enables a seamless spatial manipulation workflow. Furthermore, all UniScale techniques are strictly grounded in consistent indirect interaction, motivated by prior work~\cite{lystbaek2024hands} demonstrating its benefits for interaction coherence.

Our primary research goal is to systematically explore and analyze unimanual mapping strategies. To do so, we evaluate how different physical metaphors and clutching mechanisms affect scaling performance, physical fatigue, and user preference compared to traditional bimanual techniques. Thereby, we propose five distinct techniques, each embodying a unique scaling metaphor (Figure~\ref{fig:techniques}): (1) \textit{uniDepth} and (2) \textit{uniAngle}, which utilize non-direct mappings (i.e., physical movement amplitude determines output value, unlike the 1:1 isomorphic mapping)~\cite{mossel20133dtouch} via non-dominant hand (NDH) forward-backward translation and wrist rotation, respectively; (3) \textit{uniMicro}, a low-calorie micro-gesture controlled by NDH thumb~\cite{kin2024stmg, kim2025t2iray}; (4) \textit{uniSemi}, which provides an isomorphic scaling metaphor mapped to the NDH pinch span, toggled through hand rotation; and (5) \textit{biSemi}, an asymmetric bimanual approach where the dominant hand (DH) anchors the selection while the NDH independently controls the scale.

We conducted a user study comparing our five unimanual techniques against a baseline (\textit{biDistance}) that represents the commercially adopted bimanual symmetric pinch. We tested scaling techniques in an object resizing task in a 3D VR environment~\cite{mossel20133dtouch, frees2005precise}. Here, we asked participants to precisely match a virtual interactive cube to varying target sizes. This task examines fundamental scaling performance under two distinct interaction states, including a discrete clutching mode and a continuous clutching-free mode. The key findings reveal that the optimal clutching mode depends on the scaling gesture. Specifically, isomorphic mapping benefited from clutching mode, while rate-based mapping was better suited to a clutching-free mode. Participants preferred more frequent controls as long as they could reduce the overall physical load. Furthermore, UniScale enabled natural multitasking across complex spatial interactions by freeing the dominant hand. 

Overall, the contributions of our work are as follows: (1) We introduce UniScale, a suite of five unimanual input techniques for scaling tasks in XR that leverage indirect interaction to replace conventional two-handed scaling. (2) We identify that clutching preference varies by technique, revealing that isomorphic mappings benefit from discrete clutching, whereas rate-based mappings excel in continuous clutching-free modes. (3) We characterized suitable user contexts for each scaling technique based on user performance and subjective feedback from a VR user study.


\section{Related Work} 
\subsection{Gaze-based Indirect Interaction} 
Direct manipulation requires physical proximity to objects, limiting interaction range and causing fatigue~\cite{zeleznik2005look, yu2021gaze}. To overcome these limitations, indirect manipulation techniques decouple input from feedback, enabling interaction with distant objects without locomotion~\cite{bowman1997evaluation}. Common approaches to target selection include ray casting and gaze-assisted pointing. Ray-casting projects a virtual ray from the user's hand or controller~\cite{kim2023exploration}. Meanwhile, gaze-assisted pointing leverages eye tracking to enable rapid target acquisition~\cite{jeong2023gazehand, wang2024gazering}.

However, ray-casting suffers from precision degradation at distance: small angular changes in hand orientation produce disproportionately large cursor displacements, making fine-grained target selection challenging~\cite{bowman1998virtual, poupyrev1998egocentric}. To mitigate this, gaze has been introduced as an alternative indication modality, leveraging its speed and naturalness for rapid object pointing~\cite{pfeuffer2017gaze+}. Yet gaze alone is insufficient for precise selection because it struggles with depth disambiguation. This has motivated hybrid approaches that combine gaze for coarse target indication with hand input for fine-grained control. Gaze+Pinch demonstrated this complementarity by pairing gaze-based pointing with pinch gestures for precise manipulation~\cite{pfeuffer2017gaze+}. GazeRing further extended this by coupling gaze with pressure-sensitive ring gestures to enable fine-grained depth selection in AR~\cite{wang2024gazering}. Lystbaek et al.~\cite{lystbaek2024hands} also explored indirect NDH input for spatial referencing in bimanual tasks, finding it effective but challenging to coordinate alongside direct DH manipulation.

Subsequent work extended the gaze-based indirect interaction paradigm: GazeHand translated virtual hands to gaze-targeted locations for distant object manipulation~\cite{jeong2023gazehand}. More recent systems further refined the approach: PinchCatcher, GazeRayCursor, and PinchLens each addressed precision, multi-selection, and magnification in gaze-hand interaction~\cite{kim2025pinchcatcher, chen2023gazeraycursor, zhu2023pinchlens}. Across these systems, the pinch gesture (making thumb and index fingertip contact) has emerged as the preferred and robust input gesture, owing to its discrete activation, low muscle effort, and reliable detection in recent commercial Head-Mounted Displays (HMD) devices (e.g., Apple Vision Pro, Samsung Moohan)~\cite{apple_visionos_control, meta_mr_design_guideline}. Despite these advances, the role of Gaze+Pinch as an NDH input modality for object manipulation compared to object selection has received limited attention.

\subsection{3D Object Scaling in XR Environment}
Once a target object is successfully selected via gaze-directed pinch gestures, the interaction transitions to the scale-manipulation phase. At this stage, current controller-free 3D object manipulation techniques can be broadly categorized into bimanual and unimanual methods, each offering distinct trade-offs in expressiveness and ergonomics~\cite{piumsomboon2014grasp}. Bimanual interaction uses both hands simultaneously and can be further classified as symmetric or asymmetric. Symmetric bimanual manipulation, where both hands perform similar actions, is currently the most widely adopted technique in commercial VR/AR operating systems~\cite{meta_mr_design_guideline, ms_interaction_fundamentals, apple_visionos_control}. This technique is well-accepted by users due to its intuitive resemblance to real-world two-handed object manipulation and familiarity with the 2D multi-touch pinch-to-zoom metaphor~\cite{song2012handle}. For instance, Song et al. demonstrated this with the handle-bar metaphor, in which a virtual bar connects the user's two hands and the relative motion between them directly controls RST manipulation of the object~\cite{song2012handle}. Asymmetric bimanual interaction, where each hand performs different roles, follows Guiard's kinematic chain theory~\cite{guiard1987asymmetric}, with one hand stabilizing while the other manipulates~\cite{pierce1999voodoo, lystbaek2024hands}.

However, bimanual approaches have a significant drawback: physical fatigue caused by sustained arm elevation. The Consumed Endurance model~\cite{hincapie2014consumed} formally quantifies this phenomenon, demonstrating that cumulative muscle load during mid-air interaction directly constrains usable interaction time. Unimanual interaction offers a promising alternative to mitigate this issue. By leveraging single-handed interaction, unimanual techniques can substantially reduce the "gorilla arm" effect, thereby enabling prolonged interaction sessions without inducing discomfort. This fatigue asymmetry between bimanual and unimanual interaction has been empirically supported in UI contexts, where unimanual designs have shown reduced physical load compared to the bimanual counterparts~\cite{nyyssonen2024comparison}. 
Previous research has explored unimanual approaches across various manipulation contexts. For instance, microgesture techniques detect subtle finger movements on passive surfaces, emphasizing low-fatigue operations while providing tactile feedback~\cite{chan2016user}. Other studies have demonstrated the efficacy of unimanual manipulation for specific tasks, applying adaptive gain methods~\cite{liu2022distant} to enhance the efficiency and accuracy of translation tasks. Beyond individual hand behavior, Reynaert et al.~\cite{reynaert2023effect} showed that bimanual VR tasks exhibit strong hand synchronicity, implicating notable cognitive and motor overhead over unimanual operations. Nevertheless, unimanual scaling strategies that leverage gaze as an indirect selection modality alongside NDH input have not been systematically investigated, leaving the design space of gaze-assisted unimanual scaling unexplored.


\begin{figure*}[t]
 \centering 
 \includegraphics[width=\textwidth]{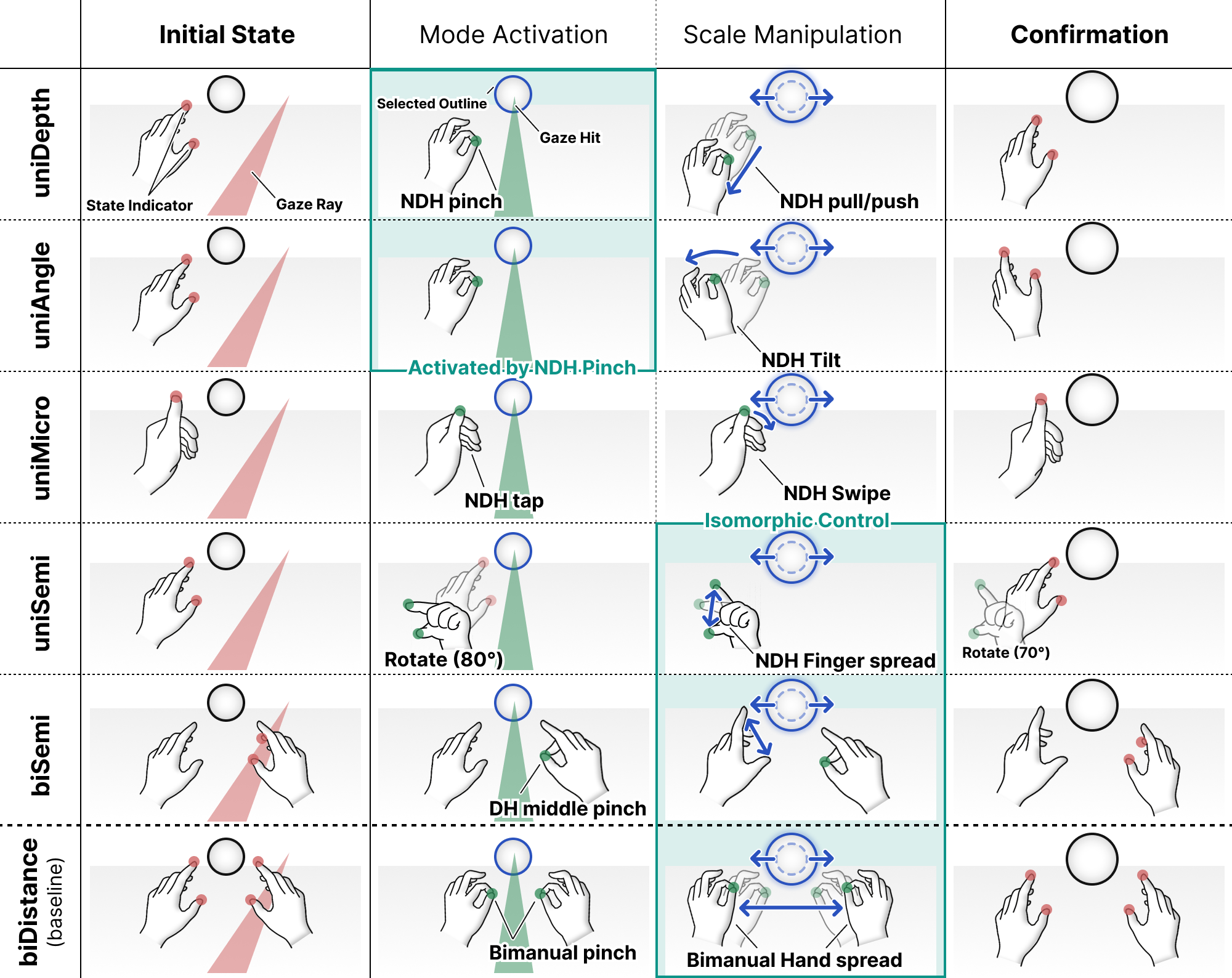}
 \caption{Overview of the proposed UniScale techniques. Interaction consists of two steps: (1) Mode activation (gaze indicates target object), where uniDepth and uniAngle use an NDH pinch, and uniMicro uses a thumbs-up micro-gesture, and (2) Scale manipulation (gaze-free), where uniSemi and biSemi adjust scale via finger span. Returning to the initial gesture terminates the interaction.}
 \label{fig:techniques}
\end{figure*}

\section{UniScale Interaction Technique Design}
We present the design of scaling techniques that seamlessly integrate natural eye movements with intuitive pinch-based manipulation within the gaze and hand gesture multimodal XR interaction paradigm~\cite{pfeuffer2017gaze+, pfeuffer2024design}. The overall manipulation process starts with a mode switch that activates the scaling state, after which the proposed gestures govern the scaling action (Figure~\ref{fig:techniques}). Here, gaze is always used to indicate the target object before mode activation, ensuring that the object is indicated before any manipulation.

In terms of scaling interaction under dominant hand-occupied conditions, we aimed to provide a unimanual alternative that delegates scaling solely to the NDH, enabling scale control without interrupting ongoing DH tasks. To structure these interactions effectively, we grounded our design on Guiard's kinematic chain model~\cite{guiard1987asymmetric,hinckley1997cooperative}, where the NDH establishes the spatial reference frame while the DH performs fine manipulation. Thereby, we assigned scaling control to the NDH and spatial translation to the DH. We further explored two operational modes across the following five techniques to examine how clutching behavior affects unimanual scaling interaction. The first mode incorporates a discrete clutching mechanism, while the second enables continuous operation without clutching.

\subsection {Scaling Gesture Design}
\subsubsection{uniDepth}
\textit{uniDepth} is a unimanual scaling technique that maps forward and backward motion to object-scale changes, which enables intuitive scaling via natural pushing and pulling~\cite{lee2025facilitating}. This technique is grounded in the concept of perceptual affordance~\cite{norman1999affordance}, whereby the perceived properties of an object suggest how it should be used. Among the gesture candidates for scaling interaction, push and pull along the depth axis are the most representative choice: the action of retracting the hand perceptually implies drawing an object closer and thus enlarging it. This alignment constitutes a strong affordance match between the motor action and its intended effect. However, as the push and pull movements are expected to easily exceed the arm reach boundary frequently, it may result in multiple clutching actions being anticipated.

\textit{uniDepth} uses an NDH pinch gesture to activate the mode, providing a clear, discrete on/off state that enables users to select objects without interfering with DH tasks. For mapping a gesture to a scaling ratio, the z-axis translation of the NDH is used. Specifically, let $z_{current}$ denote the current NDH's z-position and $z_{initial}$ the z-position at the beginning of the interaction. The depth displacement is defined as $\Delta z=z_{current}-z_{initial}$. A scale ratio is then computed as $r=1-\Delta z$, and the updated object scale is obtained by $S_{new}=S_{initial}\times r$, such that moving the hand along the z-axis continuously adjusts the object's size relative to its initial state.

\subsubsection{uniAngle}
\textit{uniAngle} is motivated from the analogous to a slider interaction, where vertical movements map linearly to a scaling output~\cite{ramos2005zliding}. A key advantage of this approach is that the required hand movement remains within the arm reach boundary, reducing the need for clutching. However, a potential drawback is that the vertical translation-to-scale mapping represents an unconventional metaphor, which may feel unnatural or awkward for users unfamiliar with this interaction model.

\textit{uniAngle} is a unimanual technique that serves as a counterpart to \textit{uniDepth}. While both share the same NDH pinch-based selection and mode activation mechanism, they differ in their spatial mapping for scaling. Whereas \textit{uniDepth} leverages depth-based (z-axis) translation, \textit{uniAngle} maps vertical (y-axis) hand movement to scale. This contrast allows us to systematically evaluate how the choice of movement axis impacts user performance and overall experience. Specifically, \textit{uniAngle} derives its scaling factor from the angular displacement of the NDH relative to its initial orientation. The angle displacement, defined as $\Delta\theta=\theta_{current}-\theta_{initial}$, is used to continuously compute the zoom ratio $r=1-\Delta\theta$.

\subsubsection{uniMicro}
\textit{uniMicro} is an unimanual technique with the thumb tap gesture~\cite{kim2025t2iray, kin2024stmg}, which is conventional for recent XR glasses. In contrast to other techniques, which require more pronounced movements, it intentionally constrains interaction to small thumb motions to investigate how it could reduce movement amplitude and affect user performance. \textit{uniMicro} uses an NDH thumb tap for scaling mode activation, which could provide quick, low-effort activation~\cite{chan2016user}. This gesture is detected when the angle formed by the thumb MCP (Metacarpophalangeal joint)-fingertip vector and the plane defined by the thumb MCP, index fingertip, and index MCP falls below $10^\circ$. To perform scaling, this technique utilizes NDH thumb swipe gestures based on angular displacement. The system evaluates the angle of the thumb's MCP-to-fingertip vector projected onto the plane formed by the thumb MCP, index MCP, and index fingertip. Using this projected angular displacement $\Delta\theta=\theta_{current}-\theta_{initial}$, the zoom ratio is computed as $r=1-\Delta\theta$, updating the final scale via $S_{new}=S_{initial}\times r$. While minimizing physical effort, a potential drawback of this method is its heavy reliance on hand-tracking stability, making interaction performance vulnerable to tracking noise or occlusion.

\subsubsection{uniSemi}
\textit{uniSemi} is a unimanual technique utilizing an isomorphic mapping between the NDH index-thumb pinch distance and object scale. This method proportionally enlarges the object as the fingers spread apart~\cite{avery2014pinch, laviola20173d}. Semi-pinch interactions are increasingly prevalent in recent works~\cite{kim2025pinchcatcher, zhu2023pinchlens} because they support fine finger manipulation while remaining consistent with the standard Gaze+Pinch paradigm. This design significantly minimizes the learning curve, as users do not need to acquire new gesture forms. However, despite the ergonomic benefits of single-handed input, it presents inherent challenges in mode switching, particularly in explicitly toggling between active manipulation and idle states.

In the current study, we used NDH wrist rotation in a semi-pinch posture for scaling mode activation for \textit{uniSemi}. Scaling mode is activated when the dot product of the palm and the user-facing direction exceeds a lock threshold of 0.75, and deactivated when it falls below an unlock threshold of 0.65, applying hysteresis to prevent unstable mode flickering near the boundary. For scaling manipulation, the NDH index-thumb pinch distance is mapped directly to the object scale ratio. As the user spreads or closes their fingers, the object scales proportionally, providing intuitive and easily understandable control~\cite{laviola20173d}.

\subsubsection{biSemi}
Lastly, \textit{biSemi} is a bimanual scaling technique counterpart to \textit{uniSemi} that employs the same isomorphic pinch mapping but explicitly separates selection and scaling between the hands~\cite{cutler1997two}. Grounded in asymmetric bimanual interaction models~\cite{guiard1987asymmetric, pierce1999voodoo}, the NDH handles scaling operation while the DH controls the mode activation. To prevent conflicts with the standard Gaze+Pinch selection paradigm~\cite{pfeuffer2017gaze+}, \textit{biSemi} is activated via a DH thumb-middle pinch, preserving the thumb-index pinch for primary operations (selection). Although this clear division of labor effectively mitigates mode-switching ambiguities, it necessitates continuous bimanual involvement.

\begin{figure}[h]
 \centering 
 \includegraphics[width=\linewidth]{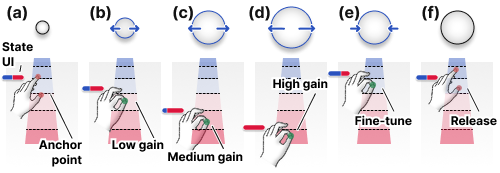}
 \caption{Illustration of clutching-free mechanism. An example of scaling up an object in \textit{uniDepth}. (a) The anchor point is established relative to the user's hand. (b)-(d) As the user performs the scale manipulation gesture (in this case, pulling), the gain increases progressively. (e) Once the desired scale is reached, the user can fine-tune by reducing the scale - without deactivating the mode - by moving the hand forward past the original anchor point. (f) The interaction ends upon release.}
 \label{fig:clutching}
\end{figure}

\subsection{Clutching Mechanism Design}
In VR environments, scaling tasks often exceed physical workspace limits, making the choice between clutching and clutching-free interactions a critical design consideration~\cite{shneiderman2010designing}. Clutching provides postural flexibility by allowing users to temporarily disengage the hand-to-object mapping and re-engage from a comfortable position~\cite{stoakley1995virtual, poupyrev1996go}. However, this segmented approach introduces physical and cognitive overhead due to repeated repositioning~\cite{argelaguet2013survey}. Conversely, clutching-free methods eliminate these discrete actions, offering smoother, continuous control~\cite{frees2007prism, buxton1990three}. Unlike translation or rotation, uniform scaling adjusts only a single scalar parameter. Thus, the required physical motion depends heavily on the specific interaction technique. Consequently, some techniques may benefit from the segmented controllability of clutching, while others prioritize clutching-free approaches to minimize physical load. Despite these distinct trade-offs, clutching mechanics in the context of scaling remain largely underexplored.

To empirically evaluate these trade-offs, we assigned each proposed technique a fixed clutching or clutching-free status, excluding dynamic switching to prevent cognitive overload~\cite{shneiderman2010designing}. In the clutching condition, users explicitly disengage (e.g., via a pinch-release gesture) at physical boundaries and re-engage to continue scaling. Conversely, the clutching-free condition produces continuous scaling as long as the activation gesture is maintained. For this clutching-free interaction, the scaling rate is proportional to hand displacement magnitude~\cite{mossel20133dtouch} and discretized into three speed tiers (slow, medium, and fast) to ensure controllability (Figure~\ref{fig:clutching}). Furthermore, a hand-mounted progress bar was displayed to provide perceptually direct feedback of both direction and magnitude~\cite{pitale2019human}. It visualizes the state by expanding the blue bar when scaling down and the red bar when scaling up.


\section{Evaluation}
\subsection{Study Design}
\subsubsection{biDistance (Baseline)}
Traditionally, scale interaction in XR has relied on manipulating the spatial distance between two input points (e.g., controllers or hands)~\cite{song2012handle, lee2025scaling}. This method is derived from the pinch-to-zoom metaphor and is widely adopted in prior work on 2D touch interactions~\cite{cutler1997two, mossel20133dtouch}. We adopt this standard approach as our baseline technique, denoted as biDistance, which dynamically maps an object's scale to the user's inter-hand distance. While we considered a UI button-based resizing method, rendering persistent UI elements for every interactable object could disrupt immersion and was therefore deemed out of scope. Furthermore, we excluded controller-based techniques because our study focuses exclusively on bare-hand interactions. Consequently, we adopted the commercially ubiquitous bimanual distance technique as our standard baseline.

Specifically, the interaction sequence operates as follows: target selection is initiated when the user gazes at an interactable object while simultaneously maintaining a stable pinch gesture with both hands. During manipulation, the object's size is directly mapped to the inter-hand distance, scaling proportionally relative to the initial distance captured at the moment of selection. The interaction terminates immediately, releasing the object, as soon as either hand exits the pinch state.

\begin{figure}[ht]
 \centering 
 \includegraphics[width=\columnwidth]{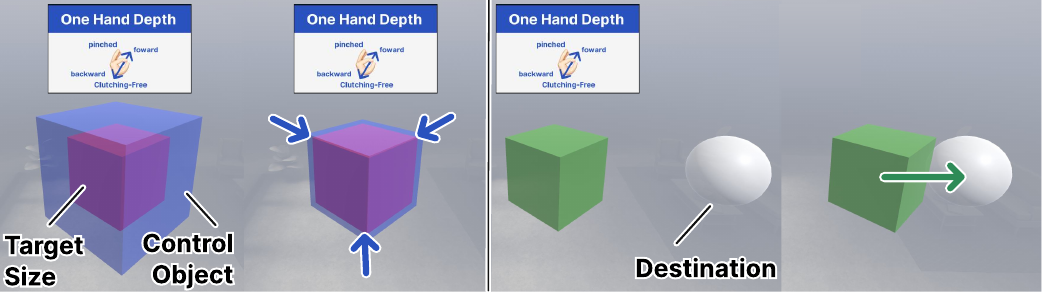}
 \caption{Overview of the user study environment, with the active condition persistently displayed in the foreground. (a) Scale task. Users manipulate the blue control cube to match the size of the red target cube. (b) Once matched, the control object turns green, and a white sphere appears to indicate the translation destination. The user then translates the green cube to the white sphere.}
 \label{fig:TaskScene}
\end{figure}

\subsubsection{Task}
We designed an object resizing task following established paradigms from prior works~\cite{mossel20133dtouch, frees2005precise}. Furthermore, we extended this to include a subsequent translation phase. This addition aimed to evaluate the technique's seamlessness with standard Gaze+Pinch~\cite{kim2025pinchcatcher, pfeuffer2024design}, specifically determining whether the scaling disrupts the fluidity of subsequent translation operations.

During the initial scaling phase, a cube with a 1 m edge length was positioned 4.5 m in front of the participant. This placement, well beyond arm's reach, ensured an indirect interaction context typical of VR environments~\cite{poupyrev1996go}. Participants were tasked with scaling the object to match one of four target edge lengths: 0.333 m, 0.577 m, 1.732 m, or 3 m. These values were selected to form equal intervals on a log scale, ensuring symmetric enlarging and shrinking conditions relative to the initial 1 m size, thereby eliminating directional bias in task difficulty. Specifically, the extreme target sizes (0.333 m and 3 m) were designed to exceed the typical physical range of a single gesture, naturally necessitating clutching behavior. In contrast, the intermediate sizes (0.577 m and 1.732 m) could be comfortably reached without clutching. The presentation order of the targets was randomized within each condition. 

A trial was completed when the participant maintained the object's size within a ±0.05 m tolerance of the target size for a continuous 500 ms dwell time, after which the object disappeared. Upon completing the scaling task, the task transitioned to the translation phase. A white target sphere appeared 4.5 m to either the left or right of the object's initial center position, and participants were required to translate and position the scaled object within this target.

\subsection{Procedure}
The experimenter obtained IRB consent forms and demographic forms from the participants and explained the tasks and scale techniques. Subsequently, participants wore the HMD and performed eye-tracking calibration. A within-subjects design was employed, and the order of techniques was randomized across participants to mitigate potential order effects and learning bias. Before starting each session, at least 5 training sessions were provided for users to learn the technique. During the trial, participants were asked to resize the interactable cube to the target cube's size using the assigned technique as accurately and as fast as possible. Participants performed the scale interaction and waited 500 ms after completing the trial. At the end of each condition, participants completed a brief questionnaire followed by a rest period, and the experimenter verbally confirmed readiness before proceeding and offered additional breaks upon request. The questionnaires consisted of NASA-TLX, SUS, and preference~\cite{hart2006nasa, brooke2013sus}. Once the questionnaire was completed, participants were given sufficient rest before proceeding to the next condition to minimize fatigue effects. After completing all conditions, participants ranked all techniques overall and participated in a brief interview about their rankings. The overall experiment lasted approximately one hour.

\subsection{Implementation}
The scale interaction techniques and the study environment were implemented using Unity 3D Engine (2022.3.39f1) and deployed on a Meta Quest Pro ($90Hz$, $106^\circ$ FoV) using its embedded eye tracker. We used Meta XR-all-in-One SDK's support for gaze and hand tracking. To reduce noise in hand tracking and gaze data, we applied the 1\texteuro~filter~\cite{casiez20121} to the raw input signals. The filter was configured with a minimum cutoff frequency of $f_{cmin}=1.0\ Hz$ and the speed coefficient of $\beta=0.07$, with a fixed derivative cutoff frequency of $1.0\ Hz$. In addition, we applied a fixation adjustment algorithm in the gaze data to minimize interaction bias triggered by gaze instability. This algorithm locks the gaze direction to the last stable orientation when the angular deviation exceeds $5^\circ$ within a $100ms$ time window.

To aid users in understanding their current interaction state and to facilitate task execution, we implemented a comprehensive visual feedback system. Following prior work~\cite{kim2025pinchcatcher}, visual cues were attached to the relevant fingertips to indicate active interaction points. For instance, as illustrated in Figure~\ref{fig:techniques}, the biDistance scaling condition featured red spheres on the index and thumb fingertips of both hands. Once the participant performed the designated interaction gesture (e.g., a pinch), these spheres transitioned to green to confirm a successful selection. Furthermore, the object's outline color provided continuous state feedback: white when gazed upon, blue during scaling, and green during translation. Finally, a spatial interface was placed at a distance in front of the user to continuously display the active experimental condition (Figure~\ref{fig:TaskScene}).

\subsection{Evaluation Metrics}
\textbf{Task Completion Time (TCT)} We defined task completion time as the interval between the initial object selection and the object release, which confirms the final size. It is the sum of the initial acquisition and fine-tuning times.
  
\textbf{Initial Acquisition Time} The initial acquisition time was measured from the initial pinch activation to the first moment the object reached the target size within the error threshold. This metric captures the ballistic phase of the task, reflecting the rapid, goal-directed adjustment toward the target.

\textbf{Fine-tuning Time} Fine-tuning time was measured as the elapsed time from when the object first reached the target size to when the user released the pinch gesture to complete the scaling task. This metric quantifies the corrective phase, where prolonged durations may indicate difficulties in precision control or stability.

\textbf{Error Rate} Error rate was calculated as the absolute difference between the final object size at pinch release and the target size, divided by the target size. This proportional metric enables fair comparison of scaling precision across trials with different target conditions.

\textbf{Attempt Count} The number of discrete activation- deactivation cycles required to complete a scaling task. This metric serves as an indicator of control stability. Higher counts indicate greater difficulty achieving the target size in a single continuous action, necessitating iterative adjustments.

\textbf{Physical Hand Movement} Hand movement was quantified as the cumulative Euclidean distance of the palm center across all frames~\cite{bashar2025early, wagner2023fitts}. For bimanual techniques, we measured both palms and summed their movements, ensuring the metric captured the total physical effort required for techniques.
  
\textbf{Questionnaire} We used the System Usability Scale (SUS)~\cite{brooke1996sus, brooke2013sus}, NASA Task Load Index~\cite{hart2006nasa}, and a single question about satisfaction on a 7-point Likert scale (0 to 6). The SUS is comprised of three positive and negative statements, with the score for the negative statement subtracted from six and subsequently summed with the positive statement scores. The NASA-TLX has six features, including mental, physical, and temporal demand, as well as perceived performance, effort, and frustration.

\textbf{Ranking \& Brief Interview} At the end of the study, participants were asked to rank all 12 conditions. Open-ended interviews were also conducted to investigate the rationale behind participants' rankings and their specific preferences.

\begin{figure*}[t]
 \centering 
 \includegraphics[width=\textwidth]{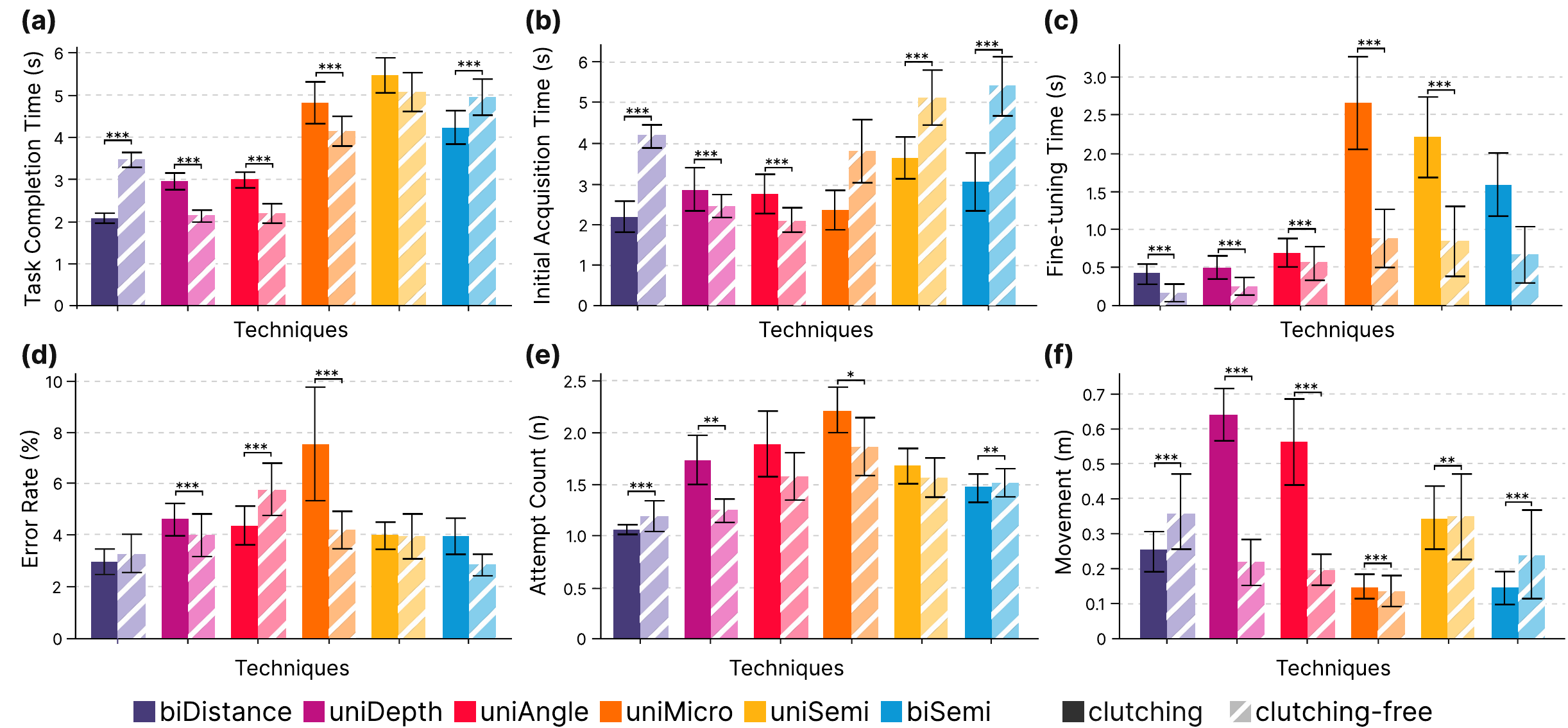}
 \caption{Results of the user study across six dependent measures. (a) Task Completion Time, (b) Initial Acquisition Time, (c) Fine-tuning Time, (d) Error Rate, (e) Attempt Count, (f) Physical Hand Movement. Statistical significance between clutching conditions is denoted by * for $p<.05$, ** for $p<.01$, and *** for $p<.001$. Error bars indicate 95\% confidence intervals.}
 \label{fig:Result}
\end{figure*}

\subsection{Participants}
We recruited 20 participants (M=25.05, SD=2.70, 15 male, 5 female) from the local university. Among the participants, all of them were right-handed, 4 wore glasses, and 2 wore contact lenses. Participants were asked to rate their prior experience with VR and hand-based interaction on a 5-point Likert scale (0–4). The participants responded to their VR experience with an average rating of 3.65 (SD=1.09). For interaction modality, they rated their experience with the hand at 2.80 (SD=1.36). All study protocols and methods were approved by the university's Institutional Review Board (IRB), and all participants provided informed consent before the experiment and were rewarded \$10 for participation.


\section{Result} 
Overall, we collected 4,800 trials total (20 participants × 2 clutching mechanisms × 6 scaling gestures  × 20 trials). Of these trials, 176 (3.67\%) were excluded from analysis. Exclusions occurred in two cases: (1) unsuccessful completion where the final size exceeded the error threshold (defined as deviation greater than 2 unity scale units from the target), and (2) outliers caused by procedural anomalies such as excessive duration, confusion, or intervention. These exclusion criteria ensured that our analysis focused on successful task executions while removing outliers that could skew performance metrics. 

Statistical analyses were performed using the Aligned Rank Transform (ART) two-way repeated measures ANOVA to accommodate non-normal data distributions~\cite{wobbrock2011aligned}. The design consisted of two within-subject factors: Scaling technique (uniDepth, uniAngle, uniMicro, uniSemi, biDistance, biSemi) and Clutching type (Clutching vs. Clutching-free). Bonferroni-corrected t-tests were used for post-hoc analysis. For NASA-TLX scores, we utilized Friedman tests, followed by pairwise Wilcoxon signed-rank tests with Bonferroni adjustments.

\subsection{Task Completion Time (Figure~\ref{fig:Result} (a))} 
Analysis of task completion time showed a significant main effect of technique ($F(5,207)=67.320, \ p<.001, \ \eta^2_p=.62$). Post-hoc pairwise comparisons showed that uniDepth ($M=2.568,\ SD=.77$) was the fastest technique, followed by uniAngle ($M=2.606,\ SD=.685$) and biDistance ($M=2.781,\ SD=.563$). These techniques were significantly faster than the remaining three techniques ($t(207)s>5.06,\ ps<.001$), although they did not differ significantly from each other ($t(207)=.547,\ p=1.000$). Conversely, uniMicro ($M=4.484,\ SD=1.597$), biSemi ($M=4.492,\ SD=1.372$), and uniSemi ($M=5.26,\ SD=1.408$) were significantly slower than the three faster techniques ($t(207)s>7.581,\ ps<.001$) but did not differ from each other ($t(207)s<2.601,\ ps>.149$). However, there was no main effect of clutching type ($F(1,207)=.337,\ p=.561,\ \eta^2_p=.001$).

There was a significant interaction effect between technique and clutching type ($F(5,207)=13.113,\ p<.001,\ \eta^2_p=.24$). uniDepth ($\Delta=.819$), uniAngle ($\Delta=.801$), and uniMicro ($\Delta=.689$) showed significantly faster with clutching-free mode ($t(207)s>2.26,\ ps<.025$), whereas biDistance ($\Delta=1.380$) and biSemi ($\Delta=.808$) were significantly slower in clutching-free mode ($t(207)s>2.67,\ ps<.009$). Finally, uniSemi showed no significant difference between clutching types ($t(207)=.63,\ p=.529$).

\subsection{Initial Acquisition Time (Figure~\ref{fig:Result} (b))}
Analysis showed a significant main effect of technique on initial acquisition time ($F(5,207)=34.2433,\ p<.001,\ \eta^2_p=.45$). Post-hoc comparisons showed that both uniAngle ($M=2.433, SD=.713$) and uniDepth ($M=2.662,\ SD=.789$) were significantly faster than baseline ($M=3.182,\ SD=.581$)($t(207)s>3.849,\ ps<.001$) whereas uniMicro ($M=3.111,\ SD=1.078$) was numerically faster but did not reach statistical significance ($t(207)=1.893,\ p=.897$). Conversely, uniSemi and biSemi were slower than baseline ($M=4.376,\ SD=1.119;\ M=4.137,\ SD=1.407$, respectively ($t(207)s>3.757,\ ps<.003$)). Moreover, there was a main effect of clutching type ($F(1,207)=96.214,\ p<.001,\ \eta^2_p=.32$): clutching status was faster ($M=2.807,\ SD=.813$) than clutching-free status ($M=3.83,\ SD=.723$).

An interaction effect was shown between technique and clutching type ($F(5,207)=24.948,\ p<.001,\ \eta^2_p=.38$). Specifically, uniAngle ($\Delta=.64,\ t(207)=6.969,\ p<.001$) and uniDepth ($\Delta=.403,\ t(207)=5.777,\ p<.001$) was significantly faster in clutching-free mode, whereas biSemi ($\Delta=2.348,\ t(207)=4.471,\ p<.001$), and biDistance ($\Delta=1.977,\ t(207)=4.475,\ p<.001$) were significantly slower in clutching-free mode. However, no significant differences were found for uniMicro ($t(207)=.765,\ p=.45$) or uniSemi ($t(207)=1.599,\ p=.11$).

\subsection{Fine-tuning Time (Figure~\ref{fig:Result} (c))}
A main effect of the technique was found ($F(5,207)=45.164,\ p<.001,\ \eta^2_p=.52$). Post-hoc pairwise comparisons showed that biDistance ($M=.299,\ SD=.214$), uniAngle ($M=.39,\ SD=.244$), and uniDepth ($M=.629,\ SD=.283$) were significantly slower than biSemi ($M=1.125,\ SD=.671$), uniSemi ($M=1.53,\ SD=.691$), and uniMicro ($M=1.734,\ SD=.835$)($t(207)s>4.042,\ ps<.001$). All pairwise comparisons showed significant differences, except for those between uniDepth and two other techniques, biDistance ($t(207)=1.369, p=1.000$) and uniAngle ($t(207)=2.706,\ p=.110$). The analysis also showed a main effect of clutching type ($F(1,207)=116.574, p<.001, \eta^2_p=.36$). Clutching-free mode ($M=.574,\ SD=.364$) was faster than the clutching condition ($M=1.325,\ SD=.355$).

An interaction effect was observed between technique and clutching status ($F(5,207)=15.591,\ p<.001,\ \eta^2_p=.27$). All techniques were faster in clutching-free mode. Notably, uniMicro ($\Delta=1.768,\ t(207)=4.983,\ p<.001$) and uniSemi ($\Delta=1.374,\ t(207)=3.021,\ p=.002$) showed the most severe degradation compared to other techniques ($t(207)s>3.659,\ ps<.001$). The only exception was biSemi, which showed no significant difference ($t(207)=1.208,\ p=.228$).

\subsection{Error Rate (Figure~\ref{fig:Result} (d))} 
There was a main effect of technique ($F(5,207)=16.970,\ p<.001,\ \eta^2_p=.29$). Post-hoc pairwise comparisons showed that biDistance ($M=3.165,\ SD=1.076$) resulted significantly lower errors than other techniques ($t(207)s>3.112, p<.031$), except for biSemi ($M=3.556,\ SD=1.298,\ t(207)=1.011,\ p=1.000$). Conversely, uniMicro ($M=5.880,\ SD=2.760$) exhibited the highest error rate, with significantly more errors than other techniques ($t(207)s>3.056,\ ps<.038$), except for uniAngle ($M=5.133,\ SD=1.584,\ t(207)=.716,\ p=1.000$). Additionally, there was a main effect in clutching status ($F(1,207)=8.871,\ p<.003,\ \eta^2_p=.04$), indicating the clutching-free mode ($M=4.583, SD=1.103$) showed a significantly higher error rate than the clutching mode ($M=4.064,\ SD=1.182$).

An interaction effect was shown between technique and clutching status ($F(5,207)=6.200,\ p<.001,\ \eta^2_p=.13$). Specifically, uniMicro exhibited the largest difference in error rate between clutching status ($\Delta=.034,\ t(207)=3.412,\ p<.001$). In contrast, uniSemi showed the smallest variation, which was not statistically significant ($\Delta=0.044,\ t(207)=.454,\ p=.650$).

\subsection{Attempt Count (Figure~\ref{fig:Result} (e))}
A main effect of technique was found ($F(5, 207)=37.010,\ p<.001,\ \eta_p^2=.47$). The Post-hoc result shows that biDistance required significantly fewer attempts ($M=1.162,\ SD=.305$) than all other techniques ($t(207)s>6.463,\ ps<.001$). Conversely, uniMicro required the highest number of attempts ($M=2.098,\ SD=.434$), significantly more than all other techniques ($t(207)s>4.219,\ ps<.001$). Moreover, the main effect of clutching status ($F(1,207)=19.134,\ p<.001,\ \eta^2_p=.08$) showed that clutching status ($M=1.732,\ SD=.271$) required more attempts in object scaling than clutching-free status ($M=1.544,\ SD=.235$).

Additionally, an interaction effect was observed ($F(5,207)=5.654,\ p<.001,\ \eta^2_p=.12$). All techniques showed fewer attempts in the clutching-free status compared to the clutching status. This difference was statistically significant for all techniques ($t(207)s>2.148,\ ps<.032$), except for uniAngle ($t(207)=.17,\ p=.865$) and uniSemi ($t(207)=.131,\ p=.896$).

\subsection{Physical Hand Movement (Figure~\ref{fig:Result} (f))}
There was a significant main effect of technique ($F(5, 207)=41.650,\ p<.001,\ \eta^2_p=.50$). Post-hoc comparisons showed that uniMicro ($M=.150,\ SD=.007$) and biSemi ($M=.197,\ SD=.169$) exhibited significantly lower hand movement than all other techniques ($t(207)s>5.945,\ ps<.001$). These were followed by biDistance ($M=.318,\ SD=.156$), uniSemi ($M=.361,\ SD=.227$), and uniAngle ($M=.395,\ SD=.17$). The greatest hand movement was observed with uniDepth ($M=.454,\ SD=.134$), which differed significantly from all other techniques ($t(207)s>4.226,\ ps<.001$, except for uniAngle ($t(207)=2.497,\ p=.119$)). There was also a significant main effect of clutching status ($F(1,207)=64.585,\ p<.001,\ \eta^2_p=.24$), indicating clutching-free status ($M=.262,\ SD=.139$) resulted in less hand movement compared to clutching status ($M=.369,\ SD=.09$).

A significant interaction effect was found ($F(5,207)=36.702,\ p<.001,\ \eta^2_p=.47$). Specifically, uniAngle and uniDepth showed substantially reduced hand movement in clutching-free status ($\Delta=.377,\ t(207)=5.987,\ p<.001;\ \Delta=.435,\ t(207)=8.784,\ p<.001$, respectively). uniMicro and uniSemi showed smaller differences ($\Delta=.001,\ t(207)=3.153,\ p=.002;\ \Delta=.001, t(207)=2.348,\ p=.019$, respectively). In contrast, biDistance and biSemi exhibited increased hand movement in clutching-free status ($\Delta=.114,\ t(207)=5.939,\ p<.001;\ \Delta=.097,\ t(207)=4.476,\ p<.001$, respectively).

\begin{figure*}[t]
 \centering 
 \includegraphics[width=\textwidth]{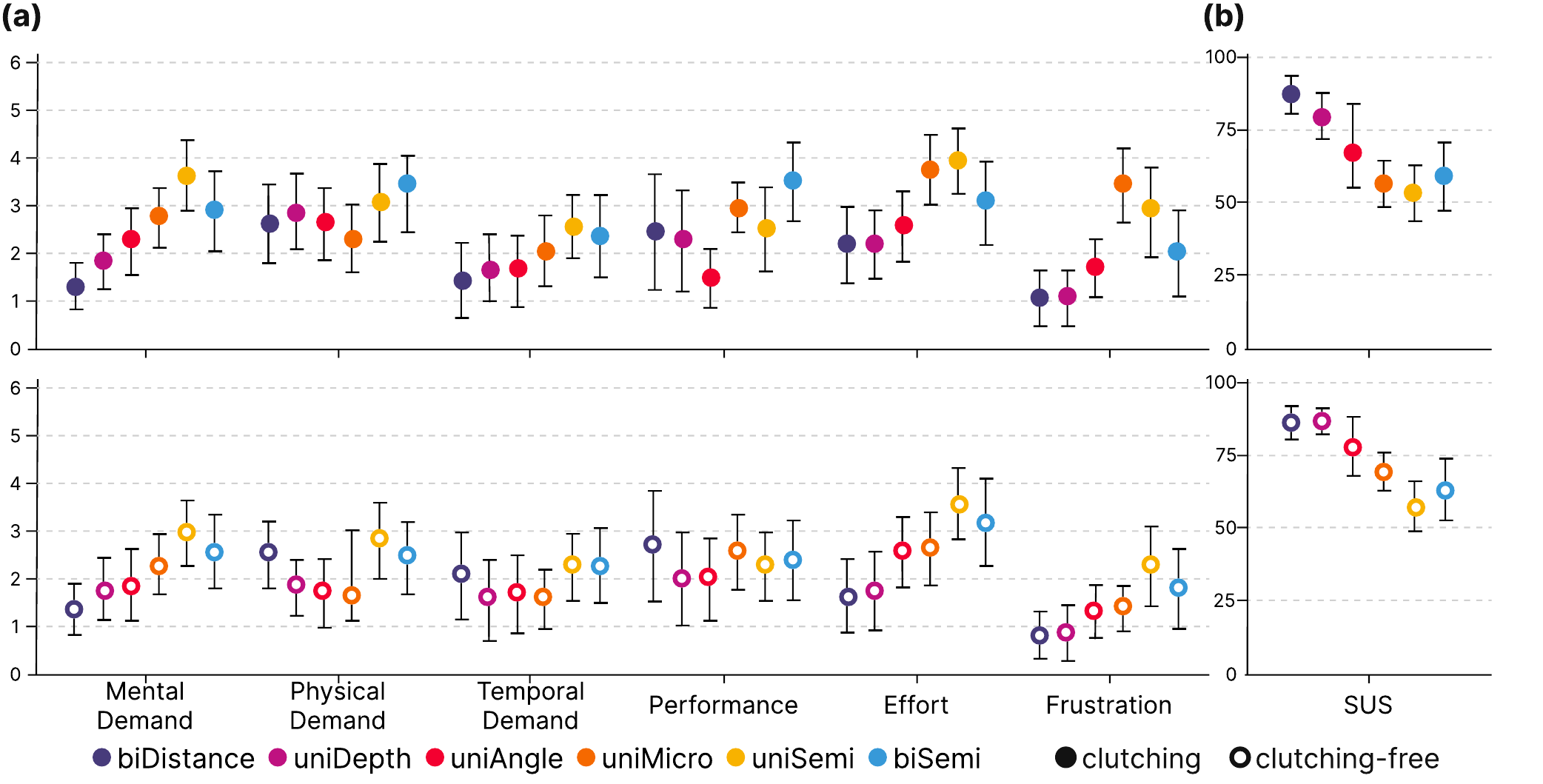}
 \caption{(a) The results of quantitative measures on the experience of using each scale technique, 7-point scale. (b) The SUS scores for each scale technique, 100-point scale. All statistical significance between clutching conditions is denoted by * for $p<.05$, ** for $p<.01$, and *** for $p<.001$. Error bars indicate 95\% confidence intervals.}
 \label{fig:questionnaire}
\end{figure*}

\subsection{Questionnaire (Figure~\ref{fig:questionnaire})} 
There was a main effect of the technique on mental demand ($\chi^2(11)=57.697,\ p<.001$), effort ($\chi^2(11)=57.393,\ p<.001$), and frustration ($\chi^2(11)=60.083,\ p<.001$) for NASA-TLX. For effort, the post-hoc result showed that uniSemi-clutching required more effort than uniDepth-clutching  ($Z=3.516,\ p=.035$). Also, biDistance-clutching-free demonstrated higher effort than uniSemi-Clutching Free ($Z=3.154,\ p=.038$). For frustration, uniMicro-clutching induced higher frustration than uniDepth-Clutching ($Z=3.516,\ p=.040$), and uniDepth-Clutching-Free ($Z=3.516,\ p=.040$). Additionally, participants experienced more frustration with uniMicro-clutching compared to uniMicro clutching-free ($Z=3.516,\ p=.038$). For mental demand, uniSemi-clutching showed higher mental demand than biDistance clutching-free ($Z=3.361,\ p=.010$).

SUS scores ranged from 52.81 (uniSemi-clutching) to 88.59 (biDistance-clutching). Eight out of twelve technique-condition combinations exceeded the acceptable usability~\cite{sauro2011practical} threshold of 68 points, with biDistance ($M_c=88.59,\ SD=8.46$;\ $M_{cf}=85.94,\ SD=11.06$) and uniDepth ($M_c=79.06,\ SD=15.33$;\ $M_{cf}=86.09,\ SD=8.61$) demonstrating excellent usability. However, uniSemi showed the lowest scores in both conditions ($M_c=52.81,\ SD=19.36$;\ $M_{cf}=57.34,\ SD=17.48$).

\begin{figure}[ht]
 \centering 
 \includegraphics[width=\columnwidth]{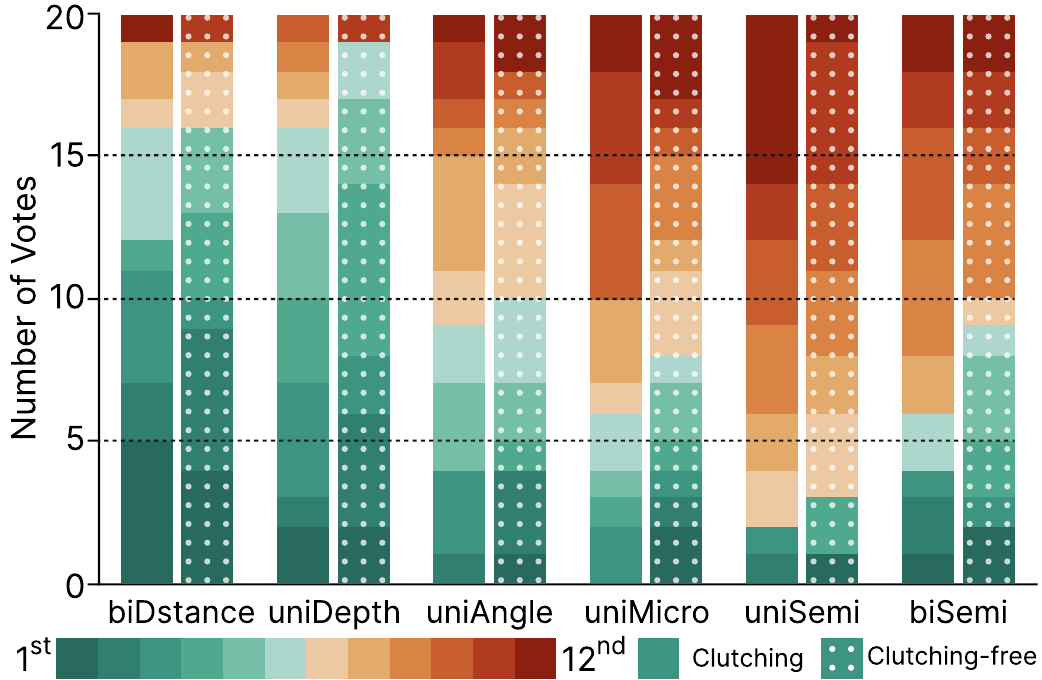}
 \caption{The result of participants' preference ranking over technique and clutching conditions.}
 \label{fig:ranking}
\end{figure}

\subsection{Ranking (Figure~\ref{fig:ranking}) and Qualitative Feedback}
Across all techniques, biDistance and uniDepth emerged as the most preferred interaction methods,  with average ranks of 3.71-4.10 and 3.86-4.71 across clutching and clutching-free conditions, respectively. Both techniques received the majority of top-3 rankings (biDistance: 11-12 out of 20, 55-60\%\; uniDepth: 7-9 out of 20, 35-45\%). uniAngle followed with an average rank of 6.29-6.76, while biSemi (average rank 7.19-7.86) and uniMicro (average rank 7.14-8.48) occupied the middle range. uniSemi was consistently ranked the least preferred technique across both conditions, recording the lowest average ranks of 8.62-9.29 and accumulating the most bottom-3 rankings.

Overall, the clutching-free condition showed improved preference rankings across most techniques, with particularly notable effects observed in uniDepth and uniMicro. uniDepth showed a substantial increase in top preference ratio from 33.33\% to 42.86\% in clutching-free mode, approaching the performance of the top-ranked biDistance, with many participants finding it intuitive and easy to learn (P2, 11, 12, 14, 16, 17). uniMicro similarly doubled its top preference ratio when clutching was removed. In contrast, biDistance maintained the highest preference across both conditions, demonstrating that its naturally familiar gesture mapping remained effective regardless of clutching status (P10, 16).

Despite these general improvements, user experiences varied considerably across techniques. uniAngle suffered from affordance issues, as the mapping between hand rotation and object scaling did not align with users' expectations (P2, 17), and uniSemi recorded the lowest overall rankings. biSemi elicited polarized opinions regarding its separated left-right hand roles, with some participants appreciating the division (P4, 14, 16, 17) while others found it problematic (P12, 15, 18, 19), though this polarization was somewhat reduced in the clutching-free condition. Overall, the clutching-free status received positive evaluations for its predictability and ease of learning (P2, 17, 20), although some participants preferred the clutching condition for its greater sense of control (P15).


\section{Discussion}
\subsection{Mapping structure as the determinant of clutching effectiveness}
Clutching demonstrated clear benefits in conditions involving isomorphic manipulation techniques (\textit{biSemi}, \textit{biDistance}). Specifically, the result showed a significantly faster task completion time and fewer attempts under the clutching condition. By establishing a direct correspondence between body state and object size, these techniques allow users to treat their physical posture as a persistent reference frame. This continuous mirroring ensures that users can seamlessly pause and resume manipulation without losing proprioceptive grounding. It enables intuitive estimation of the remaining distance to the target. This aligns with Zhai's control mapping design study~\cite{zhai1998user}, which posits that position-control interfaces are highly compatible with clutching mechanisms. However, \textit{uniSemi}, despite sharing the same finger-span scaling structure, showed no significant difference between conditions. Its threshold-based mode trigger proved unreliable in a task requiring repeated scaling, undermining the inherent advantages of its mapping structure. Overall, these findings suggest that clutching is well-suited for isomorphic mapping techniques, but only when mode activation is as simple and immediate as a discrete pinch gesture.

In contrast, clutching-free interaction proved more effective under non-isomorphic techniques (\textit{uniDepth}, \textit{uniMicro}, \textit{uniAngle}). Consistently, all three techniques showed faster task completion times and lower physical hand movements under the clutching-free condition. Unlike isomorphic techniques, these mappings lack proprioceptive grounding, which encodes the current object state to a persistent physical reference. As users interact, they naturally develop a learned sense of how much movement produces a given size change. This learned relationship acts as an implicit gain that accumulates within a continuous interaction cycle. Clutching resets this accumulated gain when re-engaging, forcing users to reinitialize their internal model. Consequently, users preferred to complete scaling within a single continuous cycle. Thus, users highly preferred clutching-free interaction for rate-based techniques. It allows users to continuously retain and build upon their learned movement-to-size mapping throughout the task without interruption.

\subsection{Gesture Affordance and Technique Preference}
\textit{uniDepth} was evaluated as the most natural interaction technique, as the affordance of pulling the hand aligns intuitively with the embodied mental model of bringing an object closer to enlarge it. This confirms that leveraging well-established perceptual affordances~\cite{norman1999affordance} can significantly reduce the cognitive gulf of execution. Such a reduction is crucial for abstract tasks like scaling that inherently lack a direct real-world physical counterpart~\cite{mendes2019survey}. In contrast, some participants found that \textit{uniAngle}, which maps wrist rotation angle to scale, felt inconsistent with the resulting scaling behavior. The rotational gesture did not intuitively convey enlargement or shrinkage, creating a perceptual mismatch between the user's action and the resulting effect. This contrast shows that aligning gestures with scaling affordances in \textit{uniDepth} can effectively replace the familiar naturalness of \textit{biDistance}.

\textit{uniMicro} presented a distinct preference profile. This technique showed the highest number of scaling attempts and a relatively high error rate among the evaluated techniques. However, its minimal hand-movement requirement led to notably positive feedback from several participants. This suggests that motor efficiency can partially compensate for a lower alignment of gesture-scale affordances. Ultimately, users' technique preferences depend not only on intuitive scaling semantics but also on the physical effort required to sustain the interaction. This preference-performance disparity stems from \textit{uniMicro}'s inherently high gain mapping, where small thumb motions within a constrained physical range produce disproportionately large scale changes. This gain enables fast initial acquisition but also amplifies unintended hand movements during fine-tuning, causing overshoot and requiring repeated corrective adjustments.

Among the semi-pinch-based techniques, \textit{biSemi} demonstrated a compelling combination of advantages. Its isomorphic finger-span mapping preserved the high spatial accuracy characteristic of position-control interfaces. At the same time, the inherently constrained range of finger motion kept physical fatigue low, achieving both precision and comfort without the typical trade-off. This validates \textit{biSemi}'s finger-spread gesture as a viable and efficient alternative to full-hand bimanual scaling~\cite{shneiderman2010designing}, mapping the user's proprioceptive sense of size directly onto the virtual object.

\subsection{Potential Implication \& Application}


\textbf{Repeated and Rapid Scaling Context} 
\textit{uniMicro} with clutching-free would show robust performance when scaling operations are repeated multiple times. This is because each additional attempt does not significantly increase the user's physical effort. Notably, \textit{uniMicro} consistently required less hand movement and overall physical demand than both the baseline and \textit{uniDepth}, proving its suitability for prolonged scaling tasks. This makes it particularly well-suited for use cases where repeated scaling is expected. For instance, exploring high-resolution images or navigating large-scale maps, where users routinely zoom in and out across many steps rather than completing the task in a single gesture.

\textbf{High-Precision Scaling Context} 
\textit{uniSemi} and \textit{biSemi} with clutching show potential, particularly in scenarios where users benefit from a strong sense of physical ownership over the scaled object. Its isomorphic mapping provides proprioceptive grounding, allowing users to feel directly coupled to the object being manipulated. These characteristics make it well-suited for tasks that demand a heightened sense of control and spatial awareness, such as precise object placement in architectural design, hands-on inspection in 3D medical visualization, or careful manipulation in XR workflows, where users need to feel as though they are physically handling the object rather than abstractly adjusting a parameter. 

\textbf{Simultaneous Workflows via Unimanual Approaches}  
The unimanual approaches could offer a distinct advantage by enabling simultaneous interactions. For example, users can concurrently scale a spatial map with one hand while annotating or routing paths with the other. This represents a meaningful shift from traditional bimanual techniques, where both hands are entirely committed to the scaling task. Furthermore, this paradigm aligns with the principles of asymmetric bimanual action~\cite{guiard1987asymmetric}. By utilizing the NDH for coarse spatial adjustments (e.g., scaling or positioning the map) and freeing the DH for fine-grained interactions (e.g., annotating). This division of labor effectively reduces the need for sequential task-switching. Consequently, this seamless parallel workflow can significantly enhance cognitive efficiency and fluid task execution in complex XR environments.



\section{Limitation \& Future Work}
Our study presents several limitations that motivate future work. First, the current experimental design evaluated scaling and translation sequentially to precisely isolate baseline performance. While our unimanual approach is available to support multitasking, future studies must evaluate it in more dynamic XR settings. Investigating the concurrent use of scaling alongside translation and rotation will be crucial to verifying its true multitasking efficiency. Next, we restricted our evaluation to scale and translation tasks. Further investigation of the simultaneous inclusion of rotation is necessary to understand user behavior in genuine multitasking scenarios. Third, magnification was bounded at 3$\times$. Real-world applications demanding extensive or repeated scaling, such as drawing tools, would likely exacerbate clutching overhead, potentially shifting user preference toward clutching-free techniques. In addition, our study scene was intentionally simplified to isolate the effect of scaling and control for confounding variables, presenting participants with a single interactable object at a time. This design may not fully reflect realistic XR applications, where users often need to manipulate multiple simultaneous objects within a cluttered or complex scene. Moreover, although concurrent multitasking was implied as a benefit of the unimanual approach, this was not directly evaluated in our study and should be empirically verified in future work. Fourth, our participant pool was limited in its diversity: all participants were right-handed, the majority were male (15 out of 20), and most reported high prior VR experience. Finally, adaptively tuning fixed implementation parameters, such as the pinch recognition threshold, based on individual hand characteristics or contextual demands could further optimize overall usability.


\section{Conclusion}
In this paper, we introduced UniScale, a suite of unimanual gesture mapping strategies designed to overcome the structural limitations and physical fatigue of bimanual scaling in Gaze+Pinch XR environments. Our evaluation of five unimanual techniques against a bimanual baseline revealed that optimal clutching depends on the mapping strategy: isomorphic mappings benefit from discrete clutching, whereas rate-based mappings excel in continuous, clutching-free modes. Furthermore, users readily accepted minor precision compromises in exchange for significantly reduced physical fatigue. By freeing the dominant hand, UniScale enables more efficient spatial multitasking. Based on these findings, we present potential implications and application scenarios where our techniques could improve the user experience of concurrent spatial manipulation and multitasking in XR environments.

\acknowledgments{This work was supported by the Institute of Information \& Communications Technology Planning \& Evaluation(IITP) grant funded by the Korea government(MSIT) (IITP-RS-2025-02214780, Generative Haptics and Fine Response Inference for Flexible Tactile Interfaces) \& (IITP-RS-2026-25507551, Development of Egocentric Data Sensing and Spatial Immersive Experience Technology)}

\bibliographystyle{abbrv-doi}
\bibliography{template}
\end{document}